\documentclass{icrc29}
\usepackage[english]{babel}
\usepackage[latin1]{inputenc}
\usepackage{times}
\usepackage[T1]{fontenc}
\usepackage{hyperref}
\usepackage{epsfig}
\usepackage{graphicx}
\usepackage{amssymb}
\hypersetup{
pdftitle={ICRC 2005 Galactic Center Paper},
pdfauthor={Antoine Letessier-Selvon, CBPF - In2p3/CNRS},
pdfborder={0 0 0},
pdfview={Fit}
}
		  
\begin{document}
\title[Galactic Center Studies] %
{Anisotropy Studies Around the Galactic Center at EeV Energies with Auger Data}

\author[Pierre Auger Collaboration]%
{The Pierre Auger Collaboration}
\presenter{Presenter: A. Letessier-Selvon (Antoine.Letessier-Selvon@@in2p3.fr),fra-letessier-selvon-A-abs1-he14-oral}

\maketitle

\begin{abstract}
The Pierre Auger Observatory data have been analyzed to search for 
excesses of events  near the direction of the galactic center in several
energy ranges around EeV energies. 
In this region the statistics 
accumulated by the Observatory are already larger than that of any 
previous experiment. Using both the data sets from the surface detector
and our hybrid data sets (events detected simultaneously by the surface 
detector and the fluorescence detector) we do not find any significant 
excess. At our present level of undestanding of the performance and properties of our detector, 
our results do not support the excesses reported by AGASA and 
SUGAR experiments. We set an upper bound on the flux of cosmic rays 
arriving within a few degrees from the galactic center in the energy 
range from 0.8-3.2~EeV. 
We also have searched for correlations of cosmic ray 
arrival directions with the galactic plane and with the super-galactic 
plane at energies in the range 1-5 EeV and above 5 EeV and have found 
no significant excess.
\end{abstract}

\section[Introduction]{Introduction}

The galactic centre (GC) region provides an attractive target for 
anisotropy studies with the Pierre Auger Observatory. On the one hand, 
there have been in the past observations by the 
AGASA experiment indicating a $4.5\sigma$ excess of cosmic rays 
(CRs) with energies in the range 1-2.5~EeV in a 
$20^\circ$ radius region centered at $(\delta,\alpha)\simeq
(-17^\circ, 280^\circ)$~\cite{ha99}.
A later search near this region with a reanalysis of SUGAR data~\cite{be00},
though with smaller statistics, failed to corroborate these findings, 
but reported a  $2.9\sigma$ excess flux of CRs with energies 0.8 <E<3.2~EeV 
in a region of $5.5^\circ$ radius centered at 
$(\delta,\alpha)=(-22^\circ,274^\circ)$.

On the other hand, since the GC harbors a very massive black hole, 
it provides a natural candidate for CR accelerator to very high energies. 
Following the recent high significance observation by HESS~\cite{ah04} 
of a TeV $\gamma$ ray source near the location of Sagittarius $A^*$, 
predictions for neutron fluxes at Auger energies have been published~\cite{NEUTRONS}.
Since neutrons emitted  by such a source would go undeflected by galactic magnetic 
fields, they should appear as a point-like source, just spread by the 
angular resolution of the experiment (the neutron decay
length becomes comparable to the distance to the GC at  EeV energies).

\section[Data Set]{Data Selection}

In this work we use Auger data from 1$^{st}$ January 2004 until 6$^{th}$ June 2005.
We use the events from the surface detector (SD)~\cite{SD} that passed the 3-fold or 
the 4-fold data acquisition triggers and satisfying our high level physics trigger (T4) 
and our quality trigger (T5)~\cite{TRIGGER}. 
The T5 selection is 
independent of energy and ensures a better quality for the event reconstruction. 
This data set has an angular resolution better than 2.2$^\circ$ for all of the 3-fold 
events (regardless of the zenith angle considered) and better than 1.7$^\circ$ for all events with
multiplicities > 3 SD stations~\cite{ANGULAR}. In all our analyses we use a zenith angle cut at 
60$^\circ$ like AGASA while SUGAR used all zenith angles.

One concern about the use of Auger data at 1~EeV given the 1.5~km spacing of our SD stations 
could be the trigger efficiency.
Various studies (using Monte Carlo simulation, our Hybrid data set, or the extrapolation of the spectral
shape) have shown that our 3-fold trigger efficiency is better than 30\% for proton induced showers and better 
than 50\% for Iron induced shower above 0.8 EeV. The 4-fold trigger reaches the same efficiency 
at about 2~EeV~\cite{ACCEPTANCE}. The efficiency ratio between Iron and proton is always
below 1.6  above 0.8~EeV and this difference can be taken into account when setting upper 
limits on point sources. 

Regarding the hybrid events (i.e. those with both Fluorescence Detectors (FD) and 
SD signal)~\cite{HYBRID}, despite the lower duty cycle of the fluorescence 
telescopes and consequently smaller statistics, they offer an excellent angular resolution
of 0.4$^\circ$~\cite{ANGULAR} over the whole energy range and a much lower trigger threshold of 0.1~EeV.

In Table~1 we show the statistics for the SD data-set 
in the various energy ranges we have used for our studies. This is the first 
search in the southern hemisphere since the SUGAR analysis and we have over 10 times 
more statistics.

\begin{table}[!t]
\label{DATA}
\caption{Events statistics}
\begin{center}
\begin{tabular}{l|c|c|c|c|c}
Data Set  & $>$0.1 EeV & [0.8-3.2] EeV & [1.0 - 5.0]EeV & $>$5.0 EeV & $>$10.0 EeV\\ \hline
SD &  122636 & 41792 & 29773 & 1359 & 387 
\end{tabular}
\end{center}
\vspace*{-0.5cm}
\end{table}

\section[Previous claims]{AGASA and SUGAR Excesses}
To estimate the coverage map, needed to construct excess and excess probability maps, we tested two different techniques.
The so-called ``shuffling'' technique 
where one uses the data to construct randomized isotropic 
data sets and, alternatively for the  SD sample, a semi-analytical method described in~\cite{COVERAGE}.
Both techniques are equivalent for studying sources because the Poisson noise in our search window dominates over the 
uncertainty of our coverage map. This report is based on the shuffling technique. 
For the hybrid sample, we have defined 24  epochs each  corresponding to a given telescope configuration. 
For the SD sample we have randomized, within 5 bands of zenith angle, the UTC hours and Julian days of the events and drawn 
the azimuth from a uniform distribution. We present in Fig.~1A the coverage map obtained from our SD sample in a 
region around the GC. Since we are well within the field of view of Auger there are no strong variation within this region.

\begin{figure}[t]
\begin{center}
\epsfig{file=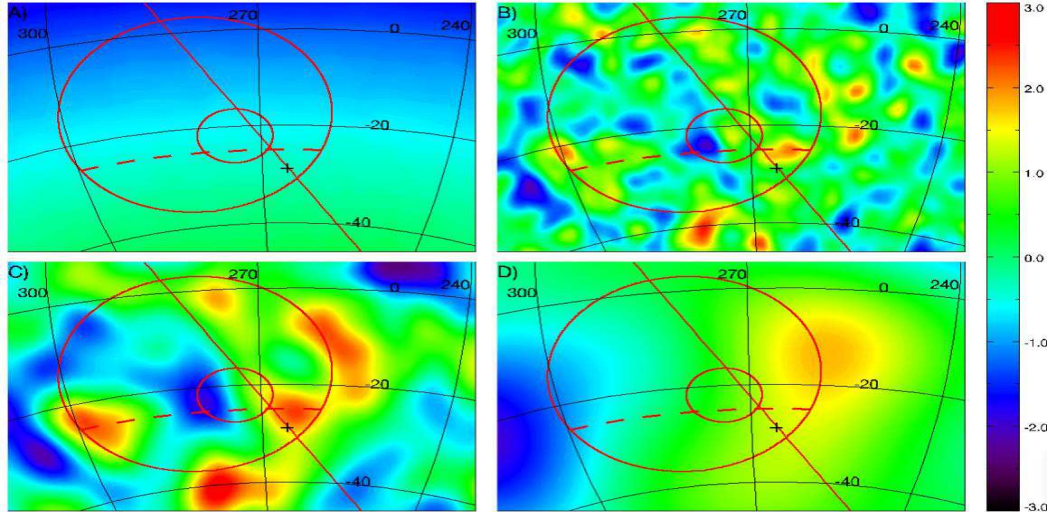,width=.9\textwidth,height=6.8cm}
\end{center}
\vspace*{-0.5cm}
\caption{Lambert projections of the galactic centre region, GC (cross), galactic plane (solid line),  regions of excess of AGASA and SUGAR (circles), AGASA f.o.v. limit (dashed line). 
A) coverage map (same color scale as the significance maps, but in a range [0-1.0]). 
B) significance map in the range [0.8-3.2]~EeV smoothed using the individual pointing resolution of the events and 
a 1.5$^\circ$ filter (Auger like excess),
C) same  smoothed at 3.7$^\circ$ (SUGAR like excess), D) in the range [1.0-2.5]~EeV smoothed at 13.3$^\circ$ (AGASA like excess).}
\label{fig:1}
\end{figure}

In Fig.1B,C,D we present the chance probability distributions (mapped to positive Gaussian significance for excesses and
negative for deficits) in the same region for various filtering and energy cuts corresponding to our various searches. 
In these map the filtering is choosen as to maximize the appearance of eventual structures at a certain scale. 
1.5$^\circ$ corresponds to our 2.2$^\circ$ angular resolution and therefore to point sources, 3.7$^\circ$ is similar 
to a 5$^\circ$ top-hat window and also  corresponds to the SUGAR excess size, 13.3$^\circ$ is similar to a 
20$^\circ$ top-hat and to the size of the excess reported by AGASA.
In these maps the chance probability distributions are consistent with those expected as a result of statistical 
fluctuations from an isotropic sky. 

Regarding the region where the AGASA excess was reported, the results from the Auger Observatory are $1155$ events observed,
and $1160.7$ expected (ratio 1.00$\pm$0.03) for the energy range [1.0-2.5]~EeV.
This is almost 3 times the event number of AGASA in this region due to the fact
that the GC lies well within the field of view of Auger while it lies outside 
the one of AGASA. These results do not support the excess observed by AGASA, and
in particular not at a level of 22\% like the one they reported which would translate into 
a 7.5$\sigma$ excess. In a worst case scenario where the source would be protons and the background much heavier (e.g. Iron), the
difference in detection efficiency of the Auger trigger at 1~EeV would reduce the sensitivity to a source excess. However,
using the Fe/proton efficiency ratio at 1~EeV ($70\%/50\%=1.44$, an upper bound in the range [1-2.5]~EeV) 
we would still expect to see a 5.2$\sigma$ event excess in our data set.

There may be systematic differences between the energy calibrations of the two experiments, 
so we have scanned the region maintaining the high/low energy ratio and moving the interval center by
$\pm40\%$. We have also enlarged the interval to [0.8--3.2]~EeV. We found no significant 
excess in any of those cases.
Regarding the excess claimed  by SUGAR, we find in their angular/energy window 
$144$ events observed, and $150.9$  expected (ratio 0.95$\pm$0.08) 
, and hence with over an order of magnitude more statistics we are not able to confirm this claim.

\section[Galactic Center]{Galactic Center}
We then searched for signals of a point-like source in the direction 
of the GC. Using a 1.5$^\circ$ Gaussian filter corresponding to the angular resolution of the SD~\cite{ANGULAR}.
In the energy range 
[0.8--3.2]~EeV, we obtain $24.3$ events observed and, $23.9$ expected (ratio 1.0$\pm$0.1).
A  95\% CL upper bound on  the number of events coming from a point source in that window is 
$n_s(95\%)= 6.3$. This bound can be translated into a 
flux upper limit if we know how many events ($n_s$) are expected for 
a given flux ($\Phi_s$) integrated in this energy range.
Since the detector efficiency is energy (and primary composition) dependent at EeV energies, the
ratio $\Phi_s/n_s$  will depend  on the spectral shape and nature of the source.
In the simplest case in which the source has a spectrum
similar to the one of the overall CR spectrum (d$N/{\rm d}E\propto E^{-3}$),
we can relate the two with $\Phi_s = n_s \Phi_{CR} 4\pi\sigma^2/n_{exp}$ 
where $\sigma$ is the size of the Gaussian filter used.
Using $\Phi_{CR}(E)= 1.5\ \xi (E/EeV)^{-3} \times 10^{-12}\, (\mbox{\rm EeV$^{-1}$\,m$^{-2}$\,s$^{-1}$\,sr$^{-1}$})$  
where $\xi \in [1,2.5]$ denotes our uncertainty on the CR flux ($\xi$ is around unity for Auger and 2.5 for AGASA), 
introducing $\varepsilon$ the Iron/proton detection  efficiency ratio ($1< \varepsilon < 1.6$ for $ E \in [0.8,3.2]$~EeV) and,
integrating in that energy range we obtain :
$$
\Phi_s  < 2.5\,\, \xi \,\, \varepsilon \times 10^{-15}\, \mbox{\rm m$^{-2}$s$^{-1}$} \mbox{~~~~@ 95\% CL.}
$$

Since the detector efficiency grows with energy, a harder source flux 
would lead to a stronger flux bound. In a worst case scenario, where both $\xi$ and $\varepsilon$ 
take their maximum value, the bound is $\Phi_s = 10.0 \times 10^{-15}\, \mbox{\rm m$^{-2}$s$^{-1}$}$, and 
still excludes the neutron source scenario suggested in~\cite{ha99,bo03} to
account for the AGASA excess,  or in~\cite{NEUTRONS} in connection with the HESS measurements.

Using the better angular resolution of the hybrid reconstruction,
we can set a point-like flux limit inside a $1^{\circ}$ radius
cone around the direction of the galactic center.
Due to the limited statistics, all hybrid events with energies above
0.1~EeV (10589) are used in this analysis, we find in this window $4$ events observed and, $3.4$ expected
, showing no significant excess. 
At 95\% CL an upper limit for a neutron source at the GC above 0.1~EeV is:
$$
\Phi_s  < 1.2\,\,\xi\times 10^{-13} \mbox{\rm m$^{-2}$s$^{-1}$} \mbox{~~~~@ 95\% CL.}
$$
 
\begin{figure}[t]
\begin{minipage}[t]{0.5\textwidth}
\begin{center}
\epsfig{file=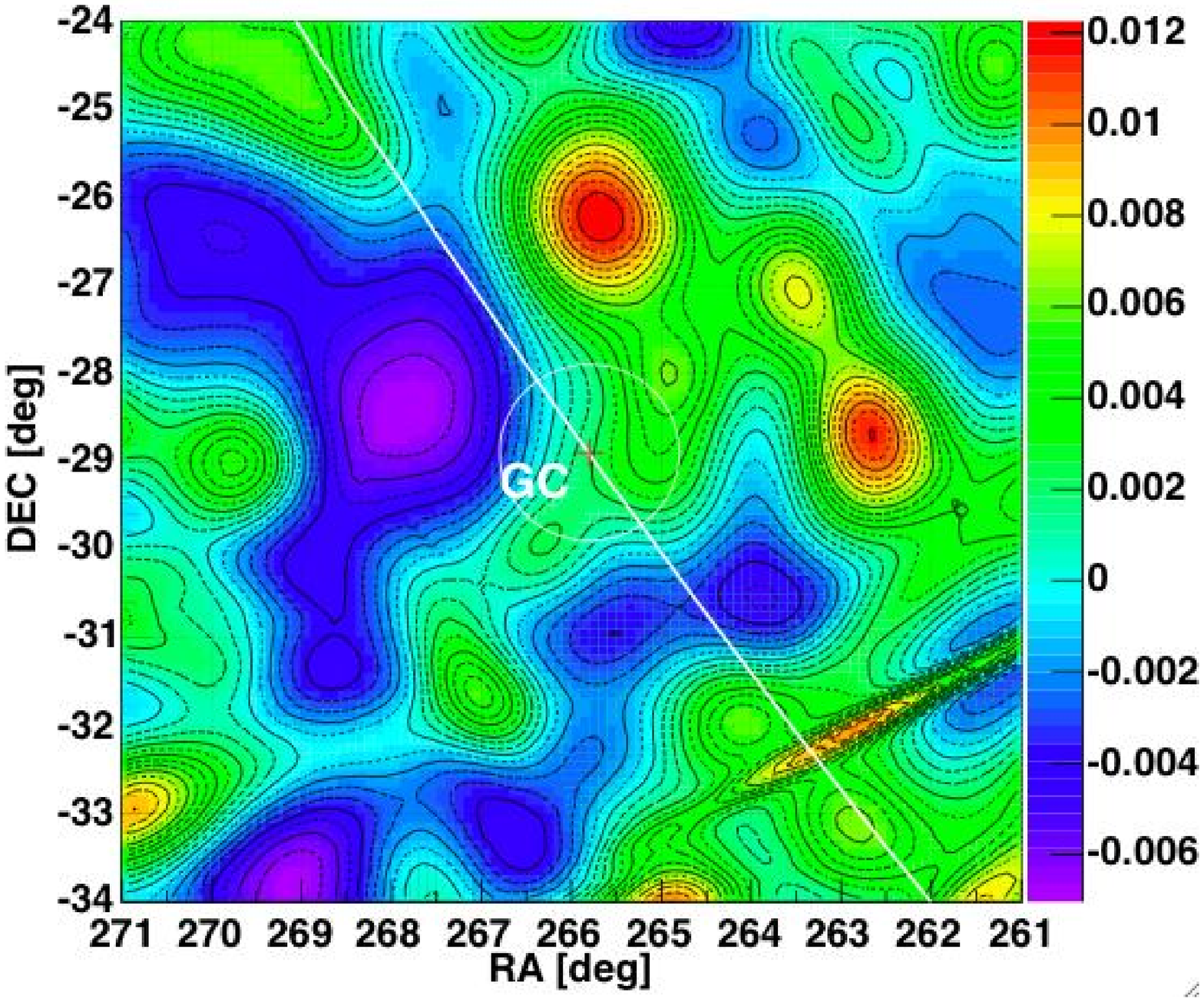,width=1.0\textwidth, height=4.8cm}  
\end{center}
\end{minipage}
\begin{minipage}[t]{0.5\textwidth}
\begin{center}
\epsfig{file=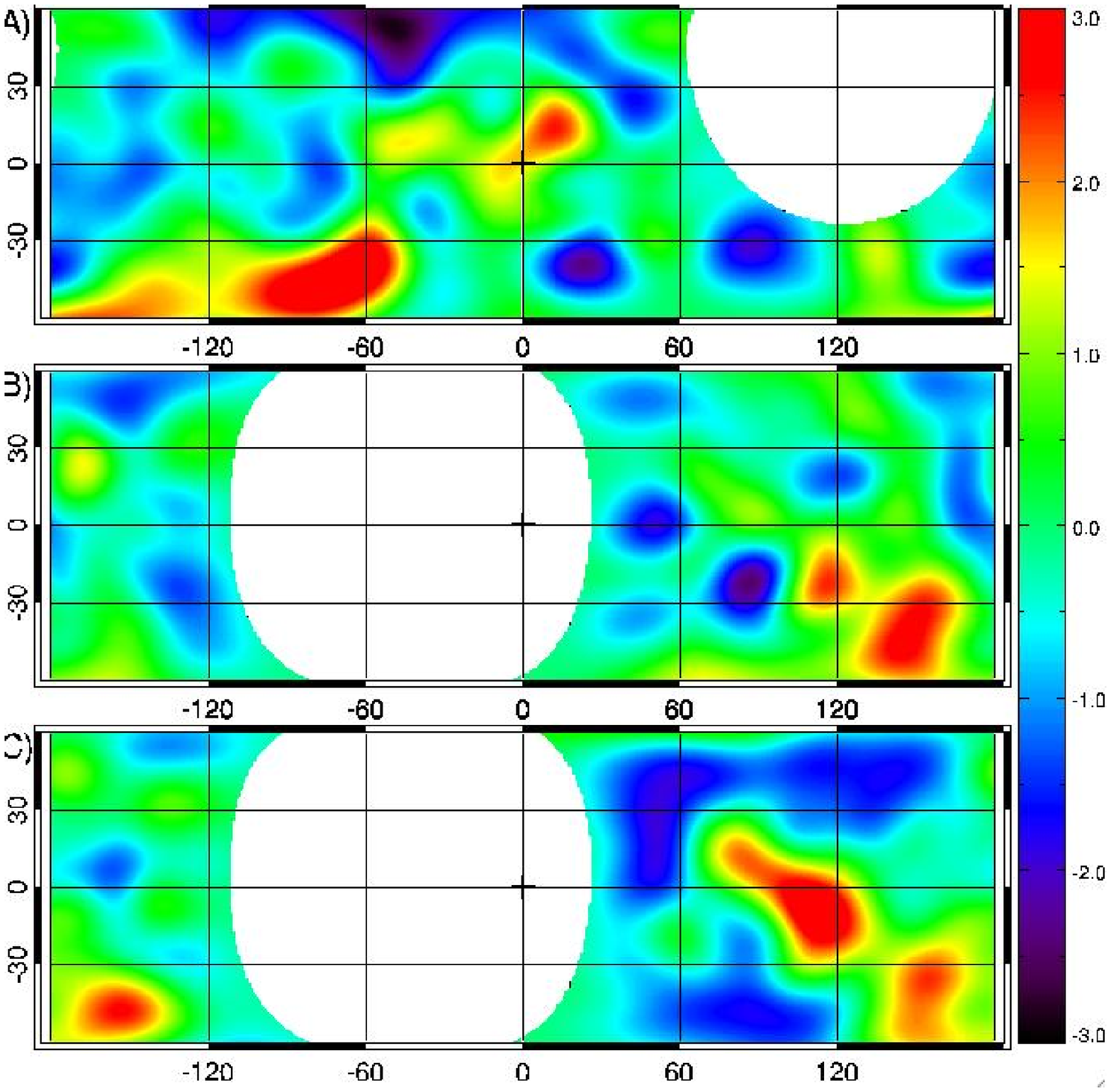,width=1.0\textwidth, height=4.8cm}  
\end{center}
\end{minipage}
\vspace*{-0.7cm}
\caption{Left, the GC region seen by the hybrid detector, the excess map is built
using the individual pointing resolution of the events.
Right, significance map from the SD data in galactic coordinates for events with 1~EeV $< E <$ 5~EeV (top),  
in super-galactic coordinates  for events with E$>$5~EeV (middle), idem for events with E$>$10~EeV (bottom).}
\label{fig:2}
\end{figure}

\vspace*{-0.5cm}
\section[Galactic and Super-Galactic Planes]{ Galactic/Super-Galactic Plane Studies}
It is expected that the origin of cosmic rays changes from galactic to extra galactic
in the 1-10 EeV range. 
We have looked for an excess of events inside a $\pm$10$^\circ$ band along the Galactic Plane in the energy range from 1~EeV to 5~EeV and along the Super Galactic Planes above 5 and 10~EeV.
We observed 5077, 229, and 68 events respectively for expectations of 5083.3, 235.6, and 67.4 (ratios 1.00$\pm$0.01, 0.97$\pm$0.07, and 1.0$\pm$0.1) showing no significant deviations. 
On Fig.~3 we show our corresponding probability  maps smoothed on a 10$^\circ$  scale.
Again all chance probabilities are consistent with isotropy.

\end{document}